\documentclass[preprint,superscriptaddress]{revtex4-1}
\pdfoutput=1 
\usepackage{txfonts}
\usepackage{CJK} 
\usepackage{float}
\usepackage{here}
\usepackage{mathrsfs}
\usepackage{here}
\usepackage{mathrsfs}
\usepackage[dvipdfmx]{graphicx}
\usepackage{color}
\begin{document}
\begin{CJK*}{}{} 
\title{Observation of evanescent spin waves in the magnetic dipole regime}
 \author{Keita~Matsumoto}
 \affiliation{Department of Physics, Tokyo Institute of Technology, Tokyo 152-8551, Japan}
 \affiliation{Department of Physics, Kyushu University, Fukuoka 819-0385, Japan}
 \author{Isao~Yoshimine}
 \affiliation{RIKEN Center for Advanced Photonics, RIKEN, Sendai 980-0845, Japan}
  \affiliation{Institute of Industrial Science, The University of Tokyo, Tokyo 153-8505, Japan}
 \author{Kosei~Himeno}
 \affiliation{Department of Physics, Kyushu University, Fukuoka 819-0385, Japan}
 \author{Tsutomu~Shimura}
  \affiliation{Institute of Industrial Science, The University of Tokyo, Tokyo 153-8505, Japan}
 \author{Takuya~Satoh}
  \email{satoh@phys.titech.ac.jp}
 \affiliation{Department of Physics, Tokyo Institute of Technology, Tokyo 152-8551, Japan}
 \affiliation{Department of Physics, Kyushu University, Fukuoka 819-0385, Japan}
 \affiliation{Institute of Industrial Science, The University of Tokyo, Tokyo 153-8505, Japan}

\begin{abstract}
We observed spin-wave transmission through an air gap that works as a prohibited region. 
The spin waves were excited by circularly polarized pump pulses via the inverse Faraday effect, and their spatial propagation was detected through the Faraday effect of probe pulses using a pump-probe imaging technique. 
The experimentally observed spin-wave transmission was reproduced using numerical calculations with a Green's function method and micromagnetic simulation. 
We found that the amplitude of the spin waves decays exponentially in the air gap, which indicates the existence of evanescent spin waves in the magnetic dipole regime. 
This finding will pave the way for controllable amplitudes and phases of spin waves propagating through an artificial magnonic crystal.
\end{abstract}

\maketitle
\end{CJK*}
\section{Introduction}
When an electromagnetic plane wave propagates from a more refractive medium towards a less refractive medium at an incidence angle exceeding the critical angle, the electromagnetic wave cannot propagate in the second medium and is totally internally reflected. Nevertheless, waves exist within the second medium, which are referred to as evanescent waves \cite{evanescentTEXT}. 
The amplitude of an electromagnetic field decreases exponentially along an interface normal and can be expressed as $\sim \exp(-Akw)$. Here, $A$ is of the order of unity when the incidence angle far exceeds the critical angle, $k$ is the wave number, and $w$ is the distance from the interface. 
This $\exp(-Akw)$ characteristic of evanescent waves should exist in every type of wave and not just electromagnetic waves.

Spin waves are propagating waves of precessing magnetization in a magnetically ordered material. 
They have been studied extensively because a spin wave can propagate in insulators with a long propagation length \cite{Kajiwara10} with dispersion that is controllable using external conditions \cite{Hurben95,Hurben96,Chumak17_magnonic_crystal}. 
The interference of multiple spin waves can be applied for the creation of novel logic gates \cite{Chumak17_magnonic_crystal}. 
In recent years, it has been reported that spin-wave transmission through artificial periodic magnetic inhomogeneities known as magnonic crystals has a spin-wave forbidden band gap in wave number space \cite{Chumak08,Vogel15}. 
Analogous to the electromagnetic case, evanescent spin waves are expected where the spin-wave propagation is prohibited in real space.
Previous research \cite{Kostylev07,Schneider10} reported the existence of a spin-wave tunneling effect through an air gap.
However, to our knowledge, evanescent spin waves with the $\exp(-Akw)$ characteristic have not been verified, because of the requirement of high temporal and spatial resolution.

An all-optical pump-probe technique with femtosecond laser pulses provides sufficient temporal resolution and yields the phase information of coherent oscillation in particular.
This technique has been widely used for spin-wave experiments \cite{Kampen02,Lenk11} in a noncontact manner. 
Spin precession can be excited nonthermally by a pump pulse via the inverse Faraday effect \cite{Kimel05}, where a pump pulse with circular polarization induces an effective magnetic field within a material. 
This effect causes no heating because the excitation process is based on impulsive stimulated Raman scattering in a nonresonant condition \cite{Kalashnikova08}.
The spatial distribution of a generated magnetic field is proportional to the intensity profile of the pump spot on the sample surface \cite{Satoh12}.
This impulsive field exerts a torque on magnetization, leading to spin precession that propagates as a spin wave out of the pump spot \cite{Satoh12,Parchenko13,Yoshimine17,Savochkin17}. 
The time-resolved imaging of spin-wave propagation has been performed using magneto-optical effects \cite{Tamaru02,Satoh12,Au13,Yoshimine14,Ogawa15,Busse15,Iihama16,Yoshimine17,Hashimoto17,Khokhlov19}. 

In this paper, we show the dynamics of spin-wave transmission through an air gap using a time-resolved pump-probe magneto-optical imaging technique.
The results are compared with numerical calculations using a Green's function method \cite{Demokritov04,Kostylev07,Schneider10} and micromagnetic simulation \cite{Vansteenkiste14} describing long-range magnetic dipole interaction.
We interpreted the transmission effect by drawing an analogy with evanescent phenomena.

\section{Method}
Our sample was an epitaxially grown single crystal of a (111)-oriented 110-$\mu$m-thick ferrimagnetic insulator, Gd$ _{3/2}$Yb$ _{1/2}$BiFe$_{5}$O$_{12}$. 
Its Curie temperature was 573 K. 
This sample is a suitable magnetic material for the optical observation of spin-wave transmission through an air gap because it shows large magneto-optical interaction and long spin-wave propagation \cite{Satoh12,Parchenko13,Yoshimine14,Yoshimine17,Chekhov18,Matsumoto18}.

\begin{figure}[htbp]
	\centering
	\includegraphics[width=8.6cm]{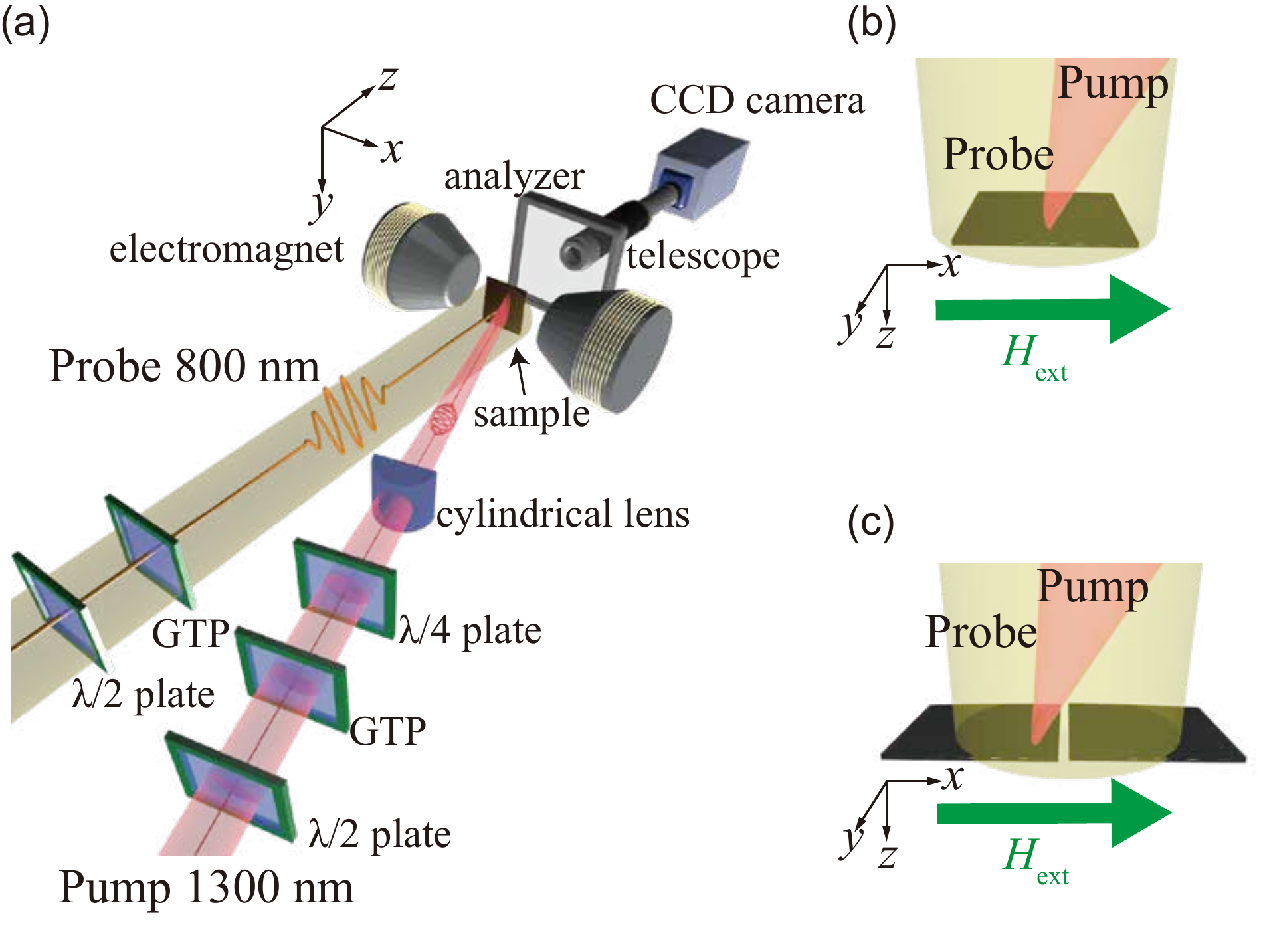}
	\caption{The experimental setup and sample geometries. (a) The optical system used in our experiments. $\lambda/2$ plate: Half-wave plate. GTP: Glan Taylor prism. $\lambda/4$ plate: Quarter-wave plate. The electromagnet applies an in-plane magnetic field, and a CCD camera detects the transmitted probe light that is guided by a telescope. Sample and pump-probe geometry: (b) Gap-free and (c) finite-gap cases.}
\end{figure}

The experimental setup is shown in Fig. 1(a). 
The light pulse for the pump-probe measurement was generated by a Ti:sapphire regenerative amplifier with a pulse duration of 120 fs. 
A circularly polarized pump light pulse with a central wavelength of 1300 nm was focused on the sample in a linear shape with a width of 20 $\mu$m using a cylindrical lens. 
This leads to the excitation of a spin wave via the inverse Faraday effect and one-dimensional propagation of the spin wave along an external magnetic field of 1000 Oe.
We utilized a linearly polarized probe pulse with a central wavelength of 800 nm.
The polarization rotation of the transmitted probe light enables us to obtain the phase and amplitude information of the spin wave via the Faraday effect, which is sensitive to the out-of-plane component of magnetization $m_{z}$.
The spatially resolved profile of the propagating spin wave was obtained by a CCD camera \cite{Yoshimine14}. 
All the measurements were performed at room temperature.

\section{Numerical calculation}
In the Green's function method, the out-of-plane component of the magnetization profile $m_z(x,k)$ at position $x$ for wave number $k$ in the presence of an air gap is calculated as \cite{Demokritov04,Kostylev07,Schneider10}
\begin{eqnarray}
m_{z}(x,k)=&\int^{\infty}_{-\infty} G(x-x',k)\left(\frac{1}{\chi}-\frac{1}{\chi_{0}}\right)m_{z}(x',k)dx'\nonumber\\
&+G(x-x_{0},k).\label{eq_green}
\end{eqnarray}
Here, $G(x,k)$ is the magnetostatic Green's function of the spin wave, $x_{0}$ is the excitation position in the left sample ($x_{0}=0$ for our calculation), and the magnetic susceptibility $\chi$ is zero in the air gap and equal to $\chi_{0}$ otherwise, where
\begin{eqnarray}
\chi_{0}= \frac{4\pi M_{\text{s}}H_{\text{ext}}}{H_{\text{ext}}(H_{\text{ext}}-H_{\text{u}})-\omega(k)^2/\gamma^2}.
\end{eqnarray}
$H_{\text{u}}$ is the uniaxial anisotropic field, $M_{\text{s}}$ is the saturation magnetization, $H_{\text{ext}}$ is the externally applied magnetic field, $\omega(k)$ is the angular frequency with dispersion of the backward volume magnetostatic wave (BVMSW) in the lowest order \cite{Damon61,Hurben95,Hurben96}, and $\gamma$ is the gyromagnetic ratio.
All the parameters for the Green's function method are described in Appendix A. 
The first term on the right-hand side of Eq. (\ref{eq_green}) expresses the dipole interaction by the precession of magnetization, whereas the second term represents the incident spin wave. 
It is assumed that the spin wave is uniformly excited along the sample thickness ($z$ direction), with its validity discussed in Appendix B. The Green's function $G(x,k)$ is described as
\begin{eqnarray}
G(x,k)=\frac{1}{2\pi}\int_{-\infty}^{\infty}\frac{\exp(-ik'x)}{W(k')-W(k)-iv}dk',
\end{eqnarray}
where $ W(k)=[\exp(-|k|L)-1]/(|k|L)$, $ L $ is the sample thickness, and $v$ is a phenomenological parameter.

Because the first term of Eq. (1) diverges in the air gap ($\chi=0$), we transformed Eq. (1) using the dipole field $h_{z}=m_{z}/\chi$ as
\begin{eqnarray}
\chi h_{z}(x,k)=&\int_{-\infty}^{\infty} G(x-x',k)\left(\frac{\chi_{0}-\chi}{\chi_{0}}\right)h_{z}(x',k)dx'\nonumber\\
&+G(x-x_{0},k).
\end{eqnarray}
This self-consistent equation takes into account multiple reflections of the dipole field in the gap.

We applied the Green's function method to the dynamics of spin-wave transmission through an air gap. 
The time-dependent magnetization $m_{z}(x,t)$ can then be represented by the superposition of $m_{z}(x,k)$ in the $k$ space \cite{Satoh12},
\begin{eqnarray}
&m_{z}(x,t)=\nonumber\\&\int \exp\left(-\frac{k^2r_{0}^2}{2}\right)\exp[i\omega(k) t]\exp[-\alpha\omega(k) t]m_{z}(x,k)dk.
\end{eqnarray}
Here, $r_{0}$ is the spatial width of the pump light and $\alpha$ is the Gilbert damping constant. 

\section{Results}
We investigated two different geometries, i.e., the gap-free case [Fig. 1(b)] and the finite-gap case [Fig. 1(c)]. 
The obtained spatiotemporal maps of spin waves represented as an out-of-plane component of magnetization $m_{z}$ with these geometries are shown in Figs. 2(a) and 2(b), respectively.

\begin{figure}[htbp]
	\centering
	\includegraphics[width=8.6cm]{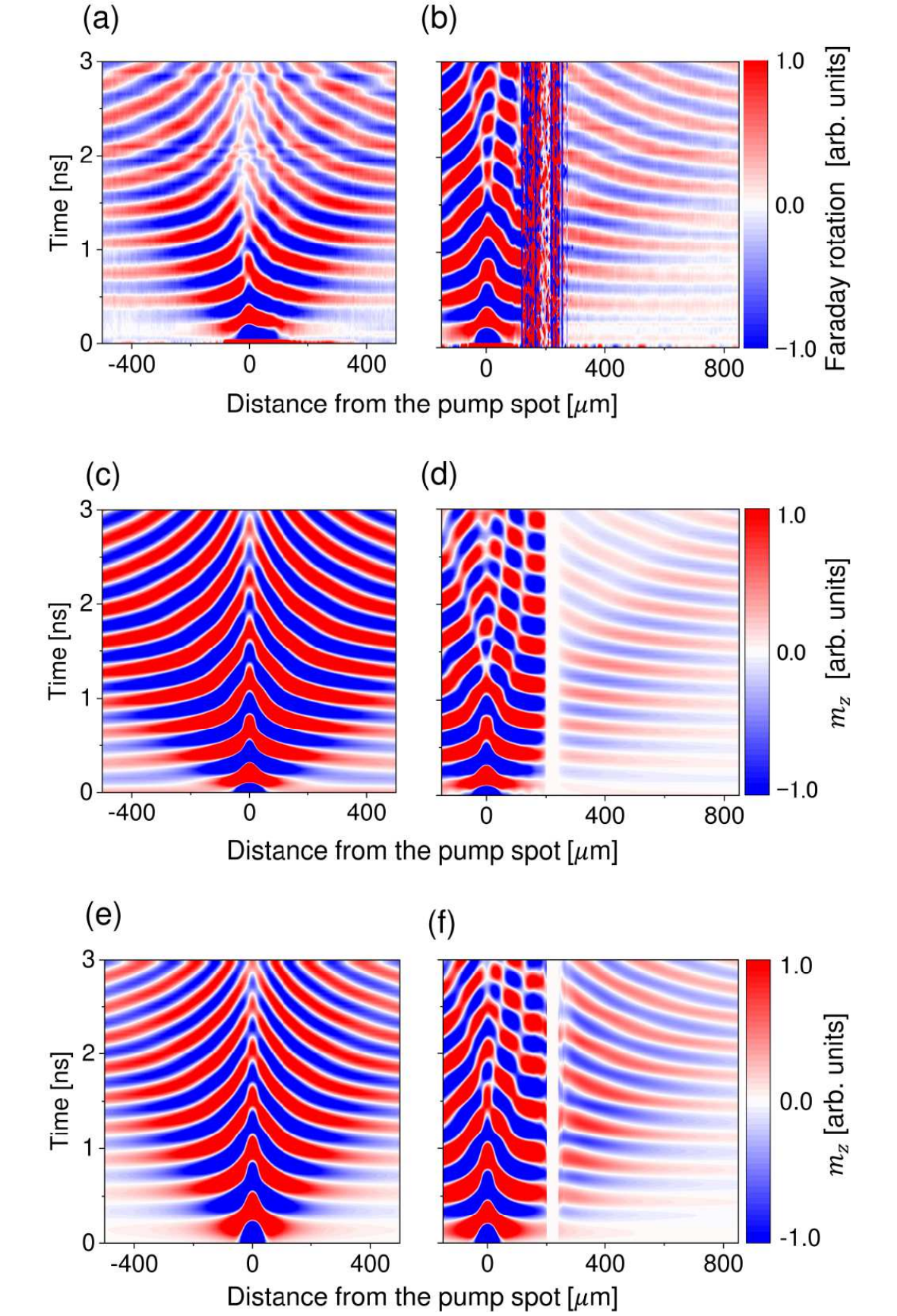}
	\caption{The experimental and calculated results of spin-wave dynamics as a function of distance from the pump spot along the $x$ axis and time. Spatiotemporal maps for the gap-free system [(a), (c), (e)] and the finite-gap system [(b), (d), (f)]; these are Faraday rotation maps obtained in the experiment [(a), (b)], the perpendicular magnetization component calculated using the Green's function method [(c), (d)], and that calculated by micromagnetic simulation [(e), (f)]. In (b), (d), and (f), the gap ranges from $x=200$ to 240~$\mu$m in the $x$ axis.}
\end{figure}

First, for the gap-free case, Fig. 2(a) indicates the excitation and propagation of spin waves with magnetic dipole characteristics---BVMSW. 
The direction of phase velocity is opposite to that of group velocity because the dispersion curve has a negative slope \cite{Satoh12}.
Figure 2(c) demonstrates the results obtained by the Green's function method. 
The agreement between Figs. 2(a) and 2(c) confirms that the observed spin waves in experiments were magnetic dipole-dominated BVMSW with the lowest order. 
In addition, the Green's function $G(x,k)$ as expressed in Eq. (3) was proved to be valid up to $k \approx 0.1$~rad$/\mu$m which was the upper limit for the excited spin wave in the experiment.

Second, the results for the finite gap with width $w=40~\mu$m, obtained by experiment and the numerical calculation by the Green's function method, are shown in Figs. 2(b) and 2(d), respectively (also see the movie in the Supplemental Material \cite{SM20}). 
The air gap is located at $x=200$--240 $\mu$m as displayed in Fig. 1(c). 
The noisy signal near the air gap in Fig. 2(b) is due to the random reflection of probe light at the sample edge.
On the left side of the pump spot ($x<0$) in Figs. 2(b) and 2(d), the propagation characteristic was almost identical to that of the
gap-free case in Figs. 2(a) and 2(c). 
On the right side of the pump spot in the left sample in Figs. 2(b) and 2(d), the spin wave exhibits a standing wave ($0<x<200~\mu$m), and its node point of interference approaches the sample edge over time. 
This is explained as follows.
The spatially focused pump pulse excites a spin-wave packet with a broadband wave number. 
Due to the dispersion of the BVMSW with negative slope, the wave with a lower wave number reaches the edge first, following which the wave with a higher wave number reaches the edge.

It is worth noting that the transmitted wave was discernible in the right sample. 
In addition, a higher transmission was observed for a lower wave number.
These experimental characteristics [Fig. 2(b)] were reproduced well by the numerical calculation [Fig. 2(d)] qualitatively and quantitatively.

To confirm the validity of the calculation by the Green's function method, we also performed a micromagnetic simulation using MUMAX$^3$ \cite{Vansteenkiste14} (for parameters, see Appendix A) as shown in Figs. 2(e) and 2(f) for the gap-free and finite-gap cases, respectively. 
The results reproduced the calculation by the Green's function method. 
Here, only the dipole interaction was taken into account and the exchange stiffness was set to zero in the micromagnetic simulation. 
The waveform exhibited a negligible change even if we set the actual exchange stiffness.
The excellent agreement among the results of the experiment, the numerical calculations of the Green's function method, and the micromagnetic simulation indicates that the transmission of the spin wave through the air gap originates from long-range magnetic dipole interaction between the left and right samples.

\section{Discussion}
The numerical calculation with the Green's function method enables us to evaluate the spin-wave transmission phenomena quantitatively. 
We calculated the spin-wave transmittance defined as $T=|m_{z, \text{t}}(x=x_{\text{far}},k)/m_{z, \text{i}}(x=x_{\text{far}},k)|$, where $m_{z, \text{i}}(x,k)$ and $ m_{z, \text{t}}(x,k)$ are the waveforms of the incident and the transmitted spin waves, respectively. 
In the calculation of the transmittance, the excitation source ($x_{0}=0$) was sufficiently far to the left of the air gap, and $x_{\text{far}}$ was sufficiently far to the right of the air gap, i.e., $x_{\text{far}}$ is of the order of centimeters, where $T$ converges to a constant value. The phase shift is defined as $\arg[m_{z, \text{t}}(x_{\text{far}},k)/m_{z, \text{i}}(x_{\text{far}},k)]$ and is the phase difference between the incident and transmitted spin waves.

\begin{figure}[htbp]
	\centering
	\includegraphics[width=8.6cm]{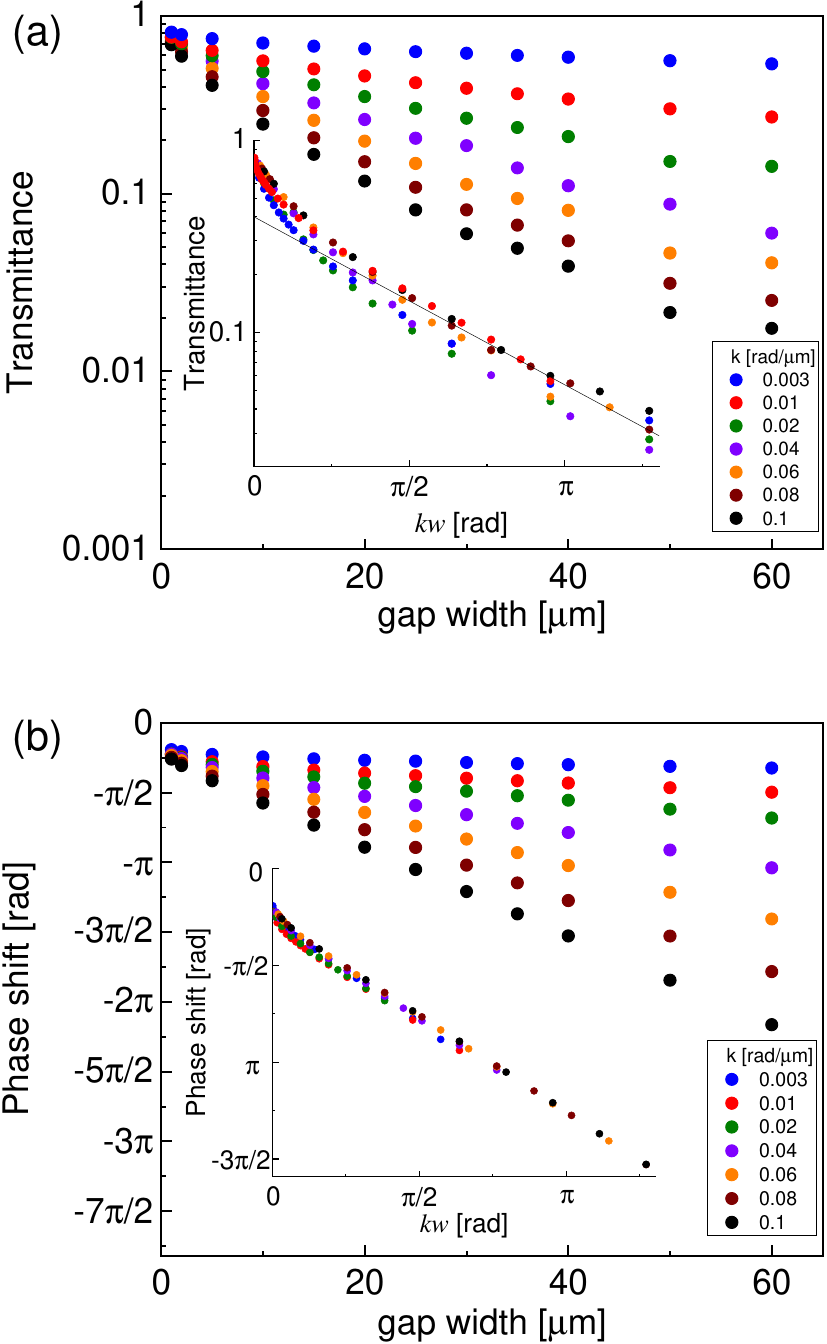}
	\caption{Calculated transmittance and phase shift of the spin wave. (a) Transmittance as a function of the gap widths $w$ for various $k$ and as a function of $kw$ in the inset. The straight line in the inset indicates $0.4 \exp(-0.64 kw)$ as a guide to the eye. (b) Phase shift as a function of the gap widths $w$ for various $k$ and as a function of $kw$ in the inset.}
\end{figure}

Figures 3(a) and 3(b) show the transmittance and phase shift of the spin wave as a function of gap width $w$ for various wave numbers $k$. 
Figure 3(a) indicates that a spin wave with a smaller wave number exhibits a higher transmission, which reproduced the experimental finding shown in Fig. 2(d). 
We replot in the insets of Figs. 3(a) and 3(b) the transmittance and the phase shift as a function of $kw$. 
Surprisingly, all curves almost coincide, forming one universal curve. 
For $kw\sim \pi$, transmittance is approximated as $\sim\exp(-Akw)$, where $A$ is of the order of unity. 
These results are reminiscent of the evanescent effect because the transmittance of electromagnetic waves can be written as $\exp(-Akw)$, where $A$ is of the order of unity.

The propagation of electromagnetic waves is generally described with the Huygens-Fresnel principle, which states that each point on an electromagnetic wave front is a source of spherical waves. 
The superposition of these waves forms the forward wavefront. 
When a plane wave is incident on an interface between two media with a finite incidence angle, these spherical waves have a phase shift along the intersection of the incident plane and the interface. 
The superposition of these phase-shifted waves then yields an evanescent wave if the incidence angle is larger than the critical angle, above which total internal reflection occurs \cite{Makris11}.
In the present case of spin waves, however, the incidence angle is zero. 
Therefore, an evanescent spin wave resulting from a phase shift along the intersection is unlikely. 
Instead, we argue that the phase shift along the propagating direction is responsible for the evanescent wave because of the long-range dipole nature of BVMSW.

\section{Conclusion}
In conclusion, we measured spatiotemporally resolved spin-wave transmission through an air gap using an all-optical pump-probe technique and found excellent agreement with the Green's function method and micromagnetic simulation.
Furthermore, we found that the transmittance calculated by the Green's function method decayed exponentially with the gap width and spin-wave wave number. 
This behavior is analogous to the evanescent phenomena of the electromagnetic wave. 
Our findings can be useful for the future development of magnonic crystals with multiple gaps (magnonic crystal). 
The observation of an evanescent spin wave will pave the way to applications in surface-sensitive devices as used in near-field spin-wave spectroscopy.

\section*{ACKNOWLEDGMENTS}
The authors thank H. Shimizu, S. Tamaru, K. Sawada, Y. Ozeki, and B. Hillebrands for valuable discussions.
We are also grateful to M. P. Kostylev for answering our inquiry.
This study was supported by Japan Society for the Promotion of Science (JSPS) (Grants No. JP15H05454, No. JP19H01828, No. JP19H05618, No. JP19J21797, No. JP19K21854, and No. JP26103004), JST-PRESTO, and JSPS Core-to-Core Program (A. Advanced Research Networks). K.M. thanks Kyushu University QR Program and Research Fellowship for Young Scientists by JSPS.

K.M. and I.Y. contributed equally to this work.

\section*{Appendix A: Simulation conditions}
We performed numerical calculations with the Green's function method and the micromagnetic simulation (MUMAX$^3$) by using the parameters listed in Table I.

\begin{table}[H]
	\centering
	\caption{Parameters for numerical calculations.}
	\begin{tabular}{ll}
		\hline
		& \qquad Parameters \\ \hline
		Uniaxial anisotropy $H_{\text{u}}$     &\qquad 600 Oe                                    \\
		Saturation magnetization $4\pi M_{\text{s}}$ &\qquad 1158 G                                       \\
		Externally applied field $H_{\text{ext}}$    &\qquad 1000 Oe                                      \\
		Gyromagnetic ratio $\gamma/$($2\pi$)            &\qquad 2.8 MHz$/$Oe \\
		Gilbert damping $\alpha$            &\qquad 0.02                                         \\
		Spatial width of pump light $r_{0}$               &\qquad 20 $\mu$m                                    \\ \hline
	\end{tabular}
\end{table}

For the Green's function method, the phenomenological parameter $ \nu =2\times10^{-5}$ was used, which is sufficiently small but still finite to ensure the convergence of the numerical calculation.

For the micromagnetic simulation, a periodic boundary condition was applied along the $y$ direction [along the sample width (see Fig. 1)] because we expect the one-dimensional propagation of the spin wave along the $x$ axis (the propagation direction). 
To simulate a spin-wave signal via the inverse Faraday effect, we consider the perturbative deviation of magnetization from the equilibrium position as $m_z(x,y)\propto\exp[-x^2/(2r_0^2)]$ \cite{Tilburg17}. 
Then, the magnetization at each cell begins precessing in accordance with the Landau-Lifshitz-Gilbert equation. 
The obtained magnetization profile was averaged along the $z$ axis for the purpose of reproducing the experimental data in the transmission geometry. 
Thus, we obtained the space- (along the $x$ axis) and time-resolved magnetization.

\section*{Appendix B: Spatial profile of the spin wave in the thickness direction}
\begin{figure}[H]
	\centering
	\includegraphics[width=8.6cm]{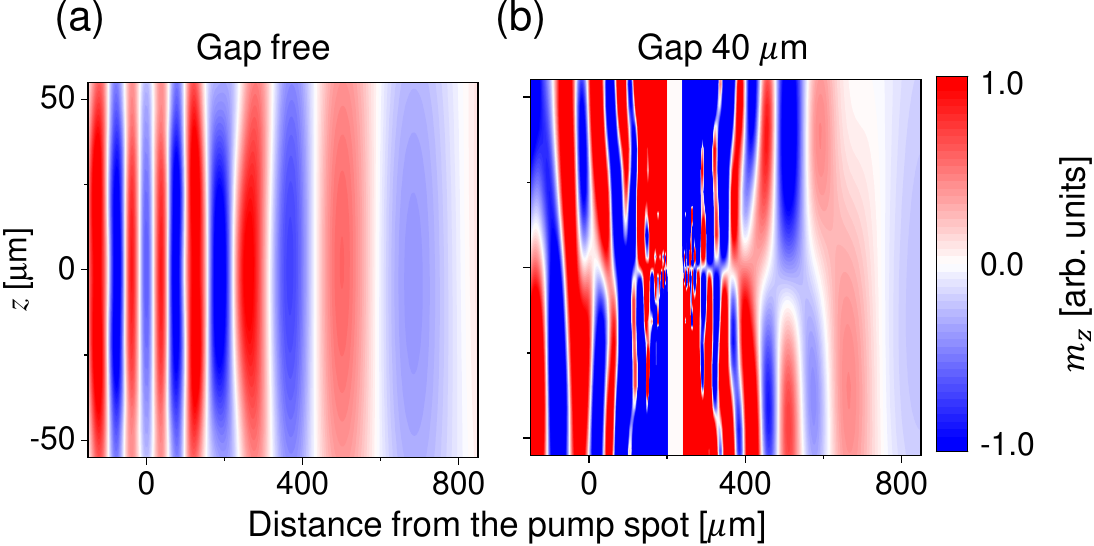}
	\caption{The spin-wave profile for (a) the gap-free case and (b) the 40-$\mu$m-gap case at 3 ns after pump excitation along the $z$ direction, which are calculated by a micromagnetic simulation.}
	\label{thickness_3ns}
\end{figure}

In the calculation using the Green's function method, we assumed that the spin wave is excited uniformly along the thickness direction ($z$). 
It is interesting to calculate the spatial profile of the spin wave in the thickness direction by a micromagnetic simulation, which is shown in Figs. \ref{thickness_3ns}(a) and  \ref{thickness_3ns}(b) at 3 ns after the pump excitation for the gap-free and 40-$\mu$m-gap cases, respectively. 
In Fig. \ref{thickness_3ns}(a), the dominant spin wave corresponds to the lowest order with an even function, and a nearly uniform profile was confirmed. 
In Fig. \ref{thickness_3ns}(b), in addition to the lowest order, the thickness profile exhibits an odd-function behavior, corresponding to the next order \cite{Hurben95}. 
The odd behavior partially results from the demagnetization near the sample edge, and it vanishes if we average along the thickness. Therefore, we can safely assume that the spin wave is excited uniformly along the thickness direction in the Green's function method.
\bibliographystyle{apsrev4-2}
\bibliography{mc}

\begin{thebibliography}{33}%
\makeatletter
\providecommand \@ifxundefined [1]{%
 \@ifx{#1\undefined}
}%
\providecommand \@ifnum [1]{%
 \ifnum #1\expandafter \@firstoftwo
 \else \expandafter \@secondoftwo
 \fi
}%
\providecommand \@ifx [1]{%
 \ifx #1\expandafter \@firstoftwo
 \else \expandafter \@secondoftwo
 \fi
}%
\providecommand \natexlab [1]{#1}%
\providecommand \enquote  [1]{``#1''}%
\providecommand \bibnamefont  [1]{#1}%
\providecommand \bibfnamefont [1]{#1}%
\providecommand \citenamefont [1]{#1}%
\providecommand \href@noop [0]{\@secondoftwo}%
\providecommand \href [0]{\begingroup \@sanitize@url \@href}%
\providecommand \@href[1]{\@@startlink{#1}\@@href}%
\providecommand \@@href[1]{\endgroup#1\@@endlink}%
\providecommand \@sanitize@url [0]{\catcode `\\12\catcode `\$12\catcode
  `\&12\catcode `\#12\catcode `\^12\catcode `\_12\catcode `\%12\relax}%
\providecommand \@@startlink[1]{}%
\providecommand \@@endlink[0]{}%
\providecommand \url  [0]{\begingroup\@sanitize@url \@url }%
\providecommand \@url [1]{\endgroup\@href {#1}{\urlprefix }}%
\providecommand \urlprefix  [0]{URL }%
\providecommand \Eprint [0]{\href }%
\providecommand \doibase [0]{https://doi.org/}%
\providecommand \selectlanguage [0]{\@gobble}%
\providecommand \bibinfo  [0]{\@secondoftwo}%
\providecommand \bibfield  [0]{\@secondoftwo}%
\providecommand \translation [1]{[#1]}%
\providecommand \BibitemOpen [0]{}%
\providecommand \bibitemStop [0]{}%
\providecommand \bibitemNoStop [0]{.\EOS\space}%
\providecommand \EOS [0]{\spacefactor3000\relax}%
\providecommand \BibitemShut  [1]{\csname bibitem#1\endcsname}%
\let\auto@bib@innerbib\@empty
\bibitem [{\citenamefont {Bertolotti}\ \emph {et~al.}(2017)\citenamefont
  {Bertolotti}, \citenamefont {Sibilia},\ and\ \citenamefont
  {Guzm\'an}}]{evanescentTEXT}%
  \BibitemOpen
  \bibfield  {author} {\bibinfo {author} {\bibfnamefont {M.}~\bibnamefont
  {Bertolotti}}, \bibinfo {author} {\bibfnamefont {C.}~\bibnamefont
  {Sibilia}},\ and\ \bibinfo {author} {\bibfnamefont {A.~M.}\ \bibnamefont
  {Guzm\'an}},\ }\href@noop {} {\emph {\bibinfo {title} {Evanescent Waves in
  Optics}}}\ (\bibinfo  {publisher} {Springer, Berlin},\ \bibinfo {year}
  {2017})\BibitemShut {NoStop}%
\bibitem [{\citenamefont {Kajiwara}\ \emph {et~al.}(2010)\citenamefont
  {Kajiwara}, \citenamefont {Harii}, \citenamefont {Takahashi}, \citenamefont
  {Ohe}, \citenamefont {Uchida}, \citenamefont {Mizuguchi}, \citenamefont
  {Umezawa}, \citenamefont {Kawai}, \citenamefont {Ando}, \citenamefont
  {Takanashi}, \citenamefont {Maekawa},\ and\ \citenamefont
  {Saitoh}}]{Kajiwara10}%
  \BibitemOpen
  \bibfield  {author} {\bibinfo {author} {\bibfnamefont {Y.}~\bibnamefont
  {Kajiwara}}, \bibinfo {author} {\bibfnamefont {K.}~\bibnamefont {Harii}},
  \bibinfo {author} {\bibfnamefont {S.}~\bibnamefont {Takahashi}}, \bibinfo
  {author} {\bibfnamefont {J.}~\bibnamefont {Ohe}}, \bibinfo {author}
  {\bibfnamefont {K.}~\bibnamefont {Uchida}}, \bibinfo {author} {\bibfnamefont
  {M.}~\bibnamefont {Mizuguchi}}, \bibinfo {author} {\bibfnamefont
  {H.}~\bibnamefont {Umezawa}}, \bibinfo {author} {\bibfnamefont
  {H.}~\bibnamefont {Kawai}}, \bibinfo {author} {\bibfnamefont
  {K.}~\bibnamefont {Ando}}, \bibinfo {author} {\bibfnamefont {K.}~\bibnamefont
  {Takanashi}}, \bibinfo {author} {\bibfnamefont {S.}~\bibnamefont {Maekawa}},\
  and\ \bibinfo {author} {\bibfnamefont {E.}~\bibnamefont {Saitoh}},\
  }\href@noop {} {\bibfield  {journal} {\bibinfo  {journal} {Nature (London)}\
  }\textbf {\bibinfo {volume} {464}},\ \bibinfo {pages} {262} (\bibinfo {year}
  {2010})}\BibitemShut {NoStop}%
\bibitem [{\citenamefont {Hurben}\ and\ \citenamefont
  {Patton}(1995)}]{Hurben95}%
  \BibitemOpen
  \bibfield  {author} {\bibinfo {author} {\bibfnamefont {M.~J.}\ \bibnamefont
  {Hurben}}\ and\ \bibinfo {author} {\bibfnamefont {C.~E.}\ \bibnamefont
  {Patton}},\ }\href@noop {} {\bibfield  {journal} {\bibinfo  {journal} {J.
  Magn. Magn. Mater.}\ }\textbf {\bibinfo {volume} {139}},\ \bibinfo {pages}
  {263} (\bibinfo {year} {1995})}\BibitemShut {NoStop}%
\bibitem [{\citenamefont {Hurben}\ and\ \citenamefont
  {Patton}(1996)}]{Hurben96}%
  \BibitemOpen
  \bibfield  {author} {\bibinfo {author} {\bibfnamefont {M.~J.}\ \bibnamefont
  {Hurben}}\ and\ \bibinfo {author} {\bibfnamefont {C.~E.}\ \bibnamefont
  {Patton}},\ }\href@noop {} {\bibfield  {journal} {\bibinfo  {journal} {J.
  Magn. Magn. Mater.}\ }\textbf {\bibinfo {volume} {163}},\ \bibinfo {pages}
  {39} (\bibinfo {year} {1996})}\BibitemShut {NoStop}%
\bibitem [{\citenamefont {Chumak}\ \emph {et~al.}(2017)\citenamefont {Chumak},
  \citenamefont {Serga},\ and\ \citenamefont
  {Hillebrands}}]{Chumak17_magnonic_crystal}%
  \BibitemOpen
  \bibfield  {author} {\bibinfo {author} {\bibfnamefont {A.~V.}\ \bibnamefont
  {Chumak}}, \bibinfo {author} {\bibfnamefont {A.~A.}\ \bibnamefont {Serga}},\
  and\ \bibinfo {author} {\bibfnamefont {B.}~\bibnamefont {Hillebrands}},\
  }\href@noop {} {\bibfield  {journal} {\bibinfo  {journal} {J. Phys. D: Appl.
  Phys.}\ }\textbf {\bibinfo {volume} {50}},\ \bibinfo {pages} {244001}
  (\bibinfo {year} {2017})}\BibitemShut {NoStop}%
\bibitem [{\citenamefont {Chumak}\ \emph {et~al.}(2008)\citenamefont {Chumak},
  \citenamefont {Serga}, \citenamefont {Hillebrands},\ and\ \citenamefont
  {Kostylev}}]{Chumak08}%
  \BibitemOpen
  \bibfield  {author} {\bibinfo {author} {\bibfnamefont {A.~V.}\ \bibnamefont
  {Chumak}}, \bibinfo {author} {\bibfnamefont {A.~A.}\ \bibnamefont {Serga}},
  \bibinfo {author} {\bibfnamefont {B.}~\bibnamefont {Hillebrands}},\ and\
  \bibinfo {author} {\bibfnamefont {M.~P.}\ \bibnamefont {Kostylev}},\
  }\href@noop {} {\bibfield  {journal} {\bibinfo  {journal} {Appl. Phys.
  Lett.}\ }\textbf {\bibinfo {volume} {93}},\ \bibinfo {pages} {022508}
  (\bibinfo {year} {2008})}\BibitemShut {NoStop}%
\bibitem [{\citenamefont {Vogel}\ \emph {et~al.}(2015)\citenamefont {Vogel},
  \citenamefont {Chumak}, \citenamefont {Waller}, \citenamefont {Langner},
  \citenamefont {Vasyuchka}, \citenamefont {Hillebrands},\ and\ \citenamefont
  {von Freymann}}]{Vogel15}%
  \BibitemOpen
  \bibfield  {author} {\bibinfo {author} {\bibfnamefont {M.}~\bibnamefont
  {Vogel}}, \bibinfo {author} {\bibfnamefont {A.~V.}\ \bibnamefont {Chumak}},
  \bibinfo {author} {\bibfnamefont {E.~H.}\ \bibnamefont {Waller}}, \bibinfo
  {author} {\bibfnamefont {T.}~\bibnamefont {Langner}}, \bibinfo {author}
  {\bibfnamefont {V.~I.}\ \bibnamefont {Vasyuchka}}, \bibinfo {author}
  {\bibfnamefont {B.}~\bibnamefont {Hillebrands}},\ and\ \bibinfo {author}
  {\bibfnamefont {G.}~\bibnamefont {von Freymann}},\ }\href@noop {} {\bibfield
  {journal} {\bibinfo  {journal} {Nat. Phys.}\ }\textbf {\bibinfo {volume}
  {11}},\ \bibinfo {pages} {487} (\bibinfo {year} {2015})}\BibitemShut
  {NoStop}%
\bibitem [{\citenamefont {Kostylev}\ \emph {et~al.}(2007)\citenamefont
  {Kostylev}, \citenamefont {Serga}, \citenamefont {Schneider}, \citenamefont
  {Neumann}, \citenamefont {Leven}, \citenamefont {Hillebrands},\ and\
  \citenamefont {Stamps}}]{Kostylev07}%
  \BibitemOpen
  \bibfield  {author} {\bibinfo {author} {\bibfnamefont {M.~P.}\ \bibnamefont
  {Kostylev}}, \bibinfo {author} {\bibfnamefont {A.~A.}\ \bibnamefont {Serga}},
  \bibinfo {author} {\bibfnamefont {T.}~\bibnamefont {Schneider}}, \bibinfo
  {author} {\bibfnamefont {T.}~\bibnamefont {Neumann}}, \bibinfo {author}
  {\bibfnamefont {B.}~\bibnamefont {Leven}}, \bibinfo {author} {\bibfnamefont
  {B.}~\bibnamefont {Hillebrands}},\ and\ \bibinfo {author} {\bibfnamefont
  {R.~L.}\ \bibnamefont {Stamps}},\ }\href@noop {} {\bibfield  {journal}
  {\bibinfo  {journal} {Phys. Rev. B}\ }\textbf {\bibinfo {volume} {76}},\
  \bibinfo {pages} {184419} (\bibinfo {year} {2007})}\BibitemShut {NoStop}%
\bibitem [{\citenamefont {Schneider}\ \emph {et~al.}(2010)\citenamefont
  {Schneider}, \citenamefont {Serga}, \citenamefont {Chumak}, \citenamefont
  {Hillebrands}, \citenamefont {Stamps},\ and\ \citenamefont
  {Kostylev}}]{Schneider10}%
  \BibitemOpen
  \bibfield  {author} {\bibinfo {author} {\bibfnamefont {T.}~\bibnamefont
  {Schneider}}, \bibinfo {author} {\bibfnamefont {A.~A.}\ \bibnamefont
  {Serga}}, \bibinfo {author} {\bibfnamefont {A.~V.}\ \bibnamefont {Chumak}},
  \bibinfo {author} {\bibfnamefont {B.}~\bibnamefont {Hillebrands}}, \bibinfo
  {author} {\bibfnamefont {R.~L.}\ \bibnamefont {Stamps}},\ and\ \bibinfo
  {author} {\bibfnamefont {M.~P.}\ \bibnamefont {Kostylev}},\ }\href@noop {}
  {\bibfield  {journal} {\bibinfo  {journal} {Europhys. Lett.}\ }\textbf
  {\bibinfo {volume} {90}},\ \bibinfo {pages} {27003} (\bibinfo {year}
  {2010})}\BibitemShut {NoStop}%
\bibitem [{\citenamefont {van Kampen}\ \emph {et~al.}(2002)\citenamefont {van
  Kampen}, \citenamefont {Jozsa}, \citenamefont {Kohlhepp}, \citenamefont
  {LeClair}, \citenamefont {Lagae}, \citenamefont {de~Jonge},\ and\
  \citenamefont {Koopmans}}]{Kampen02}%
  \BibitemOpen
  \bibfield  {author} {\bibinfo {author} {\bibfnamefont {M.}~\bibnamefont {van
  Kampen}}, \bibinfo {author} {\bibfnamefont {C.}~\bibnamefont {Jozsa}},
  \bibinfo {author} {\bibfnamefont {J.~T.}\ \bibnamefont {Kohlhepp}}, \bibinfo
  {author} {\bibfnamefont {P.}~\bibnamefont {LeClair}}, \bibinfo {author}
  {\bibfnamefont {L.}~\bibnamefont {Lagae}}, \bibinfo {author} {\bibfnamefont
  {W.~J.~M.}\ \bibnamefont {de~Jonge}},\ and\ \bibinfo {author} {\bibfnamefont
  {B.}~\bibnamefont {Koopmans}},\ }\href@noop {} {\bibfield  {journal}
  {\bibinfo  {journal} {Phys. Rev. Lett.}\ }\textbf {\bibinfo {volume} {88}},\
  \bibinfo {pages} {227201} (\bibinfo {year} {2002})}\BibitemShut {NoStop}%
\bibitem [{\citenamefont {Lenk}\ \emph {et~al.}(2011)\citenamefont {Lenk},
  \citenamefont {Ulrichs}, \citenamefont {Garbs},\ and\ \citenamefont
  {M{\"u}nzenberg}}]{Lenk11}%
  \BibitemOpen
  \bibfield  {author} {\bibinfo {author} {\bibfnamefont {B.}~\bibnamefont
  {Lenk}}, \bibinfo {author} {\bibfnamefont {H.}~\bibnamefont {Ulrichs}},
  \bibinfo {author} {\bibfnamefont {F.}~\bibnamefont {Garbs}},\ and\ \bibinfo
  {author} {\bibfnamefont {M.}~\bibnamefont {M{\"u}nzenberg}},\ }\href@noop {}
  {\bibfield  {journal} {\bibinfo  {journal} {Phys. Rep.}\ }\textbf {\bibinfo
  {volume} {507}},\ \bibinfo {pages} {107} (\bibinfo {year}
  {2011})}\BibitemShut {NoStop}%
\bibitem [{\citenamefont {Kimel}\ \emph {et~al.}(2005)\citenamefont {Kimel},
  \citenamefont {Kirilyuk}, \citenamefont {Usachev}, \citenamefont {Pisarev},
  \citenamefont {Balbashov},\ and\ \citenamefont {{Th. Rasing}}}]{Kimel05}%
  \BibitemOpen
  \bibfield  {author} {\bibinfo {author} {\bibfnamefont {A.~V.}\ \bibnamefont
  {Kimel}}, \bibinfo {author} {\bibfnamefont {A.}~\bibnamefont {Kirilyuk}},
  \bibinfo {author} {\bibfnamefont {P.~A.}\ \bibnamefont {Usachev}}, \bibinfo
  {author} {\bibfnamefont {R.~V.}\ \bibnamefont {Pisarev}}, \bibinfo {author}
  {\bibfnamefont {A.~M.}\ \bibnamefont {Balbashov}},\ and\ \bibinfo {author}
  {\bibnamefont {{Th. Rasing}}},\ }\href@noop {} {\bibfield  {journal}
  {\bibinfo  {journal} {Nature (London)}\ }\textbf {\bibinfo {volume} {435}},\
  \bibinfo {pages} {655} (\bibinfo {year} {2005})}\BibitemShut {NoStop}%
\bibitem [{\citenamefont {Kalashnikova}\ \emph {et~al.}(2008)\citenamefont
  {Kalashnikova}, \citenamefont {Kimel}, \citenamefont {Pisarev}, \citenamefont
  {Gridnev}, \citenamefont {Usachev}, \citenamefont {Kirilyuk},\ and\
  \citenamefont {{Th. Rasing}}}]{Kalashnikova08}%
  \BibitemOpen
  \bibfield  {author} {\bibinfo {author} {\bibfnamefont {A.~M.}\ \bibnamefont
  {Kalashnikova}}, \bibinfo {author} {\bibfnamefont {A.~V.}\ \bibnamefont
  {Kimel}}, \bibinfo {author} {\bibfnamefont {R.~V.}\ \bibnamefont {Pisarev}},
  \bibinfo {author} {\bibfnamefont {V.~N.}\ \bibnamefont {Gridnev}}, \bibinfo
  {author} {\bibfnamefont {P.~A.}\ \bibnamefont {Usachev}}, \bibinfo {author}
  {\bibfnamefont {A.}~\bibnamefont {Kirilyuk}},\ and\ \bibinfo {author}
  {\bibnamefont {{Th. Rasing}}},\ }\href@noop {} {\bibfield  {journal}
  {\bibinfo  {journal} {Phys. Rev. B}\ }\textbf {\bibinfo {volume} {78}},\
  \bibinfo {pages} {104301} (\bibinfo {year} {2008})}\BibitemShut {NoStop}%
\bibitem [{\citenamefont {Satoh}\ \emph {et~al.}(2012)\citenamefont {Satoh},
  \citenamefont {Terui}, \citenamefont {Moriya}, \citenamefont {Ivanov},
  \citenamefont {Ando}, \citenamefont {Saitoh}, \citenamefont {Shimura},\ and\
  \citenamefont {Kuroda}}]{Satoh12}%
  \BibitemOpen
  \bibfield  {author} {\bibinfo {author} {\bibfnamefont {T.}~\bibnamefont
  {Satoh}}, \bibinfo {author} {\bibfnamefont {Y.}~\bibnamefont {Terui}},
  \bibinfo {author} {\bibfnamefont {R.}~\bibnamefont {Moriya}}, \bibinfo
  {author} {\bibfnamefont {B.~A.}\ \bibnamefont {Ivanov}}, \bibinfo {author}
  {\bibfnamefont {K.}~\bibnamefont {Ando}}, \bibinfo {author} {\bibfnamefont
  {E.}~\bibnamefont {Saitoh}}, \bibinfo {author} {\bibfnamefont
  {T.}~\bibnamefont {Shimura}},\ and\ \bibinfo {author} {\bibfnamefont
  {K.}~\bibnamefont {Kuroda}},\ }\href@noop {} {\bibfield  {journal} {\bibinfo
  {journal} {Nat. Photonics}\ }\textbf {\bibinfo {volume} {6}},\ \bibinfo
  {pages} {662} (\bibinfo {year} {2012})}\BibitemShut {NoStop}%
\bibitem [{\citenamefont {Parchenko}\ \emph {et~al.}(2013)\citenamefont
  {Parchenko}, \citenamefont {Stupakiewicz}, \citenamefont {Yoshimine},
  \citenamefont {Satoh},\ and\ \citenamefont {Maziewski}}]{Parchenko13}%
  \BibitemOpen
  \bibfield  {author} {\bibinfo {author} {\bibfnamefont {S.}~\bibnamefont
  {Parchenko}}, \bibinfo {author} {\bibfnamefont {A.}~\bibnamefont
  {Stupakiewicz}}, \bibinfo {author} {\bibfnamefont {I.}~\bibnamefont
  {Yoshimine}}, \bibinfo {author} {\bibfnamefont {T.}~\bibnamefont {Satoh}},\
  and\ \bibinfo {author} {\bibfnamefont {A.}~\bibnamefont {Maziewski}},\
  }\href@noop {} {\bibfield  {journal} {\bibinfo  {journal} {Appl. Phys.
  Lett.}\ }\textbf {\bibinfo {volume} {103}},\ \bibinfo {pages} {172402}
  (\bibinfo {year} {2013})}\BibitemShut {NoStop}%
\bibitem [{\citenamefont {Yoshimine}\ \emph {et~al.}(2017)\citenamefont
  {Yoshimine}, \citenamefont {Tanaka}, \citenamefont {Shimura},\ and\
  \citenamefont {Satoh}}]{Yoshimine17}%
  \BibitemOpen
  \bibfield  {author} {\bibinfo {author} {\bibfnamefont {I.}~\bibnamefont
  {Yoshimine}}, \bibinfo {author} {\bibfnamefont {Y.~Y.}\ \bibnamefont
  {Tanaka}}, \bibinfo {author} {\bibfnamefont {T.}~\bibnamefont {Shimura}},\
  and\ \bibinfo {author} {\bibfnamefont {T.}~\bibnamefont {Satoh}},\
  }\href@noop {} {\bibfield  {journal} {\bibinfo  {journal} {Europhys. Lett.}\
  }\textbf {\bibinfo {volume} {117}},\ \bibinfo {pages} {67001} (\bibinfo
  {year} {2017})}\BibitemShut {NoStop}%
\bibitem [{\citenamefont {Savochkin}\ \emph {et~al.}(2017)\citenamefont
  {Savochkin}, \citenamefont {J{\"a}ckl}, \citenamefont {Belotelov},
  \citenamefont {Akimov}, \citenamefont {Kozhaev}, \citenamefont {Sylgacheva},
  \citenamefont {Chernov}, \citenamefont {Shaposhnikov}, \citenamefont
  {Prokopov}, \citenamefont {Berzhansky}, \citenamefont {Yakovlev},
  \citenamefont {Zvezdin},\ and\ \citenamefont {Bayer}}]{Savochkin17}%
  \BibitemOpen
  \bibfield  {author} {\bibinfo {author} {\bibfnamefont {I.~V.}\ \bibnamefont
  {Savochkin}}, \bibinfo {author} {\bibfnamefont {M.}~\bibnamefont
  {J{\"a}ckl}}, \bibinfo {author} {\bibfnamefont {V.~I.}\ \bibnamefont
  {Belotelov}}, \bibinfo {author} {\bibfnamefont {I.~A.}\ \bibnamefont
  {Akimov}}, \bibinfo {author} {\bibfnamefont {M.~A.}\ \bibnamefont {Kozhaev}},
  \bibinfo {author} {\bibfnamefont {D.~A.}\ \bibnamefont {Sylgacheva}},
  \bibinfo {author} {\bibfnamefont {A.~I.}\ \bibnamefont {Chernov}}, \bibinfo
  {author} {\bibfnamefont {A.~N.}\ \bibnamefont {Shaposhnikov}}, \bibinfo
  {author} {\bibfnamefont {A.~R.}\ \bibnamefont {Prokopov}}, \bibinfo {author}
  {\bibfnamefont {V.~N.}\ \bibnamefont {Berzhansky}}, \bibinfo {author}
  {\bibfnamefont {D.~R.}\ \bibnamefont {Yakovlev}}, \bibinfo {author}
  {\bibfnamefont {A.~K.}\ \bibnamefont {Zvezdin}},\ and\ \bibinfo {author}
  {\bibfnamefont {M.}~\bibnamefont {Bayer}},\ }\href@noop {} {\bibfield
  {journal} {\bibinfo  {journal} {Sci. Rep.}\ }\textbf {\bibinfo {volume}
  {7}},\ \bibinfo {pages} {5668} (\bibinfo {year} {2017})}\BibitemShut
  {NoStop}%
\bibitem [{\citenamefont {Tamaru}\ \emph {et~al.}(2002)\citenamefont {Tamaru},
  \citenamefont {Bain}, \citenamefont {van~de Veerdonk}, \citenamefont
  {Crawford}, \citenamefont {Covington},\ and\ \citenamefont
  {Kryder}}]{Tamaru02}%
  \BibitemOpen
  \bibfield  {author} {\bibinfo {author} {\bibfnamefont {S.}~\bibnamefont
  {Tamaru}}, \bibinfo {author} {\bibfnamefont {J.~A.}\ \bibnamefont {Bain}},
  \bibinfo {author} {\bibfnamefont {R.~J.~M.}\ \bibnamefont {van~de Veerdonk}},
  \bibinfo {author} {\bibfnamefont {T.~M.}\ \bibnamefont {Crawford}}, \bibinfo
  {author} {\bibfnamefont {M.}~\bibnamefont {Covington}},\ and\ \bibinfo
  {author} {\bibfnamefont {M.~H.}\ \bibnamefont {Kryder}},\ }\href@noop {}
  {\bibfield  {journal} {\bibinfo  {journal} {J. Appl. Phys.}\ }\textbf
  {\bibinfo {volume} {91}},\ \bibinfo {pages} {8034} (\bibinfo {year}
  {2002})}\BibitemShut {NoStop}%
\bibitem [{\citenamefont {Au}\ \emph {et~al.}(2013)\citenamefont {Au},
  \citenamefont {Dvornik}, \citenamefont {Davison}, \citenamefont {Ahmad},
  \citenamefont {Keatley}, \citenamefont {Vansteenkiste}, \citenamefont
  {Van~Waeyenberge},\ and\ \citenamefont {Kruglyak}}]{Au13}%
  \BibitemOpen
  \bibfield  {author} {\bibinfo {author} {\bibfnamefont {Y.}~\bibnamefont
  {Au}}, \bibinfo {author} {\bibfnamefont {M.}~\bibnamefont {Dvornik}},
  \bibinfo {author} {\bibfnamefont {T.}~\bibnamefont {Davison}}, \bibinfo
  {author} {\bibfnamefont {E.}~\bibnamefont {Ahmad}}, \bibinfo {author}
  {\bibfnamefont {P.~S.}\ \bibnamefont {Keatley}}, \bibinfo {author}
  {\bibfnamefont {A.}~\bibnamefont {Vansteenkiste}}, \bibinfo {author}
  {\bibfnamefont {B.}~\bibnamefont {Van~Waeyenberge}},\ and\ \bibinfo {author}
  {\bibfnamefont {V.~V.}\ \bibnamefont {Kruglyak}},\ }\href@noop {} {\bibfield
  {journal} {\bibinfo  {journal} {Phys. Rev. Lett.}\ }\textbf {\bibinfo
  {volume} {110}},\ \bibinfo {pages} {097201} (\bibinfo {year}
  {2013})}\BibitemShut {NoStop}%
\bibitem [{\citenamefont {Yoshimine}\ \emph {et~al.}(2014)\citenamefont
  {Yoshimine}, \citenamefont {Satoh}, \citenamefont {Iida}, \citenamefont
  {Stupakiewicz}, \citenamefont {Maziewski},\ and\ \citenamefont
  {Shimura}}]{Yoshimine14}%
  \BibitemOpen
  \bibfield  {author} {\bibinfo {author} {\bibfnamefont {I.}~\bibnamefont
  {Yoshimine}}, \bibinfo {author} {\bibfnamefont {T.}~\bibnamefont {Satoh}},
  \bibinfo {author} {\bibfnamefont {R.}~\bibnamefont {Iida}}, \bibinfo {author}
  {\bibfnamefont {A.}~\bibnamefont {Stupakiewicz}}, \bibinfo {author}
  {\bibfnamefont {A.}~\bibnamefont {Maziewski}},\ and\ \bibinfo {author}
  {\bibfnamefont {T.}~\bibnamefont {Shimura}},\ }\href@noop {} {\bibfield
  {journal} {\bibinfo  {journal} {J. Appl. Phys.}\ }\textbf {\bibinfo {volume}
  {116}},\ \bibinfo {pages} {043907} (\bibinfo {year} {2014})}\BibitemShut
  {NoStop}%
\bibitem [{\citenamefont {Ogawa}\ \emph {et~al.}(2015)\citenamefont {Ogawa},
  \citenamefont {Koshibae}, \citenamefont {Beekman}, \citenamefont {Nagaosa},
  \citenamefont {Kubota}, \citenamefont {Kawasaki},\ and\ \citenamefont
  {Tokura}}]{Ogawa15}%
  \BibitemOpen
  \bibfield  {author} {\bibinfo {author} {\bibfnamefont {N.}~\bibnamefont
  {Ogawa}}, \bibinfo {author} {\bibfnamefont {W.}~\bibnamefont {Koshibae}},
  \bibinfo {author} {\bibfnamefont {A.~J.}\ \bibnamefont {Beekman}}, \bibinfo
  {author} {\bibfnamefont {N.}~\bibnamefont {Nagaosa}}, \bibinfo {author}
  {\bibfnamefont {M.}~\bibnamefont {Kubota}}, \bibinfo {author} {\bibfnamefont
  {M.}~\bibnamefont {Kawasaki}},\ and\ \bibinfo {author} {\bibfnamefont
  {Y.}~\bibnamefont {Tokura}},\ }\href@noop {} {\bibfield  {journal} {\bibinfo
  {journal} {Proc. Natl. Acad. Sci. U.S.A.}\ }\textbf {\bibinfo {volume}
  {112}},\ \bibinfo {pages} {8977} (\bibinfo {year} {2015})}\BibitemShut
  {NoStop}%
\bibitem [{\citenamefont {Busse}\ \emph {et~al.}(2015)\citenamefont {Busse},
  \citenamefont {Mansurova}, \citenamefont {Lenk}, \citenamefont {von~der
  Ehe},\ and\ \citenamefont {M{\"u}nzenberg}}]{Busse15}%
  \BibitemOpen
  \bibfield  {author} {\bibinfo {author} {\bibfnamefont {F.}~\bibnamefont
  {Busse}}, \bibinfo {author} {\bibfnamefont {M.}~\bibnamefont {Mansurova}},
  \bibinfo {author} {\bibfnamefont {B.}~\bibnamefont {Lenk}}, \bibinfo {author}
  {\bibfnamefont {M.}~\bibnamefont {von~der Ehe}},\ and\ \bibinfo {author}
  {\bibfnamefont {M.}~\bibnamefont {M{\"u}nzenberg}},\ }\href@noop {}
  {\bibfield  {journal} {\bibinfo  {journal} {Sci. Rep.}\ }\textbf {\bibinfo
  {volume} {5}},\ \bibinfo {pages} {12824} (\bibinfo {year}
  {2015})}\BibitemShut {NoStop}%
\bibitem [{\citenamefont {Iihama}\ \emph {et~al.}(2016)\citenamefont {Iihama},
  \citenamefont {Sasaki}, \citenamefont {Sugihara}, \citenamefont {Kamimaki},
  \citenamefont {Ando},\ and\ \citenamefont {Mizukami}}]{Iihama16}%
  \BibitemOpen
  \bibfield  {author} {\bibinfo {author} {\bibfnamefont {S.}~\bibnamefont
  {Iihama}}, \bibinfo {author} {\bibfnamefont {Y.}~\bibnamefont {Sasaki}},
  \bibinfo {author} {\bibfnamefont {A.}~\bibnamefont {Sugihara}}, \bibinfo
  {author} {\bibfnamefont {A.}~\bibnamefont {Kamimaki}}, \bibinfo {author}
  {\bibfnamefont {Y.}~\bibnamefont {Ando}},\ and\ \bibinfo {author}
  {\bibfnamefont {S.}~\bibnamefont {Mizukami}},\ }\href@noop {} {\bibfield
  {journal} {\bibinfo  {journal} {Phys. Rev. B}\ }\textbf {\bibinfo {volume}
  {94}},\ \bibinfo {pages} {020401} (\bibinfo {year} {2016})}\BibitemShut
  {NoStop}%
\bibitem [{\citenamefont {Hashimoto}\ \emph {et~al.}(2017)\citenamefont
  {Hashimoto}, \citenamefont {Daimon}, \citenamefont {Iguchi}, \citenamefont
  {Oikawa}, \citenamefont {Shen}, \citenamefont {Sato}, \citenamefont
  {Bossini}, \citenamefont {Tabuchi}, \citenamefont {Satoh}, \citenamefont
  {Hillebrands}, \citenamefont {Bauer}, \citenamefont {Johansen}, \citenamefont
  {Kirilyuk}, \citenamefont {{Th. Rasing}},\ and\ \citenamefont
  {Saitoh}}]{Hashimoto17}%
  \BibitemOpen
  \bibfield  {author} {\bibinfo {author} {\bibfnamefont {Y.}~\bibnamefont
  {Hashimoto}}, \bibinfo {author} {\bibfnamefont {S.}~\bibnamefont {Daimon}},
  \bibinfo {author} {\bibfnamefont {R.}~\bibnamefont {Iguchi}}, \bibinfo
  {author} {\bibfnamefont {Y.}~\bibnamefont {Oikawa}}, \bibinfo {author}
  {\bibfnamefont {K.}~\bibnamefont {Shen}}, \bibinfo {author} {\bibfnamefont
  {K.}~\bibnamefont {Sato}}, \bibinfo {author} {\bibfnamefont {D.}~\bibnamefont
  {Bossini}}, \bibinfo {author} {\bibfnamefont {Y.}~\bibnamefont {Tabuchi}},
  \bibinfo {author} {\bibfnamefont {T.}~\bibnamefont {Satoh}}, \bibinfo
  {author} {\bibfnamefont {B.}~\bibnamefont {Hillebrands}}, \bibinfo {author}
  {\bibfnamefont {G.~E.~W.}\ \bibnamefont {Bauer}}, \bibinfo {author}
  {\bibfnamefont {T.~H.}\ \bibnamefont {Johansen}}, \bibinfo {author}
  {\bibfnamefont {A.}~\bibnamefont {Kirilyuk}}, \bibinfo {author} {\bibnamefont
  {{Th. Rasing}}},\ and\ \bibinfo {author} {\bibfnamefont {E.}~\bibnamefont
  {Saitoh}},\ }\href@noop {} {\bibfield  {journal} {\bibinfo  {journal} {Nat.
  Commun.}\ }\textbf {\bibinfo {volume} {8}},\ \bibinfo {pages} {15859}
  (\bibinfo {year} {2017})}\BibitemShut {NoStop}%
\bibitem [{\citenamefont {{N. E. Khokhlov}}\ \emph {et~al.}(2019)\citenamefont
  {{N. E. Khokhlov}}, \citenamefont {{P. I. Gerevenkov}}, \citenamefont {{L. A.
  Shelukhin}}, \citenamefont {{A. V. Azovtsev}}, \citenamefont {{N. A.
  Pertsev}}, \citenamefont {Wang}, \citenamefont {{A. W. Rushforth}},
  \citenamefont {{A. V. Scherbakov}},\ and\ \citenamefont {{A. M.
  Kalashnikova}}}]{Khokhlov19}%
  \BibitemOpen
  \bibfield  {author} {\bibinfo {author} {\bibnamefont {{N. E. Khokhlov}}},
  \bibinfo {author} {\bibnamefont {{P. I. Gerevenkov}}}, \bibinfo {author}
  {\bibnamefont {{L. A. Shelukhin}}}, \bibinfo {author} {\bibnamefont {{A. V.
  Azovtsev}}}, \bibinfo {author} {\bibnamefont {{N. A. Pertsev}}}, \bibinfo
  {author} {\bibfnamefont {M.}~\bibnamefont {Wang}}, \bibinfo {author}
  {\bibnamefont {{A. W. Rushforth}}}, \bibinfo {author} {\bibnamefont {{A. V.
  Scherbakov}}},\ and\ \bibinfo {author} {\bibnamefont {{A. M.
  Kalashnikova}}},\ }\href@noop {} {\bibfield  {journal} {\bibinfo  {journal}
  {Phys. Rev. Applied}\ }\textbf {\bibinfo {volume} {12}},\ \bibinfo {pages}
  {044044} (\bibinfo {year} {2019})}\BibitemShut {NoStop}%
\bibitem [{\citenamefont {Demokritov}\ \emph {et~al.}(2004)\citenamefont
  {Demokritov}, \citenamefont {Serga}, \citenamefont {Andr\'e}, \citenamefont
  {Demidov}, \citenamefont {Kostylev}, \citenamefont {Hillebrands},\ and\
  \citenamefont {Slavin}}]{Demokritov04}%
  \BibitemOpen
  \bibfield  {author} {\bibinfo {author} {\bibfnamefont {S.~O.}\ \bibnamefont
  {Demokritov}}, \bibinfo {author} {\bibfnamefont {A.~A.}\ \bibnamefont
  {Serga}}, \bibinfo {author} {\bibfnamefont {A.}~\bibnamefont {Andr\'e}},
  \bibinfo {author} {\bibfnamefont {V.~E.}\ \bibnamefont {Demidov}}, \bibinfo
  {author} {\bibfnamefont {M.~P.}\ \bibnamefont {Kostylev}}, \bibinfo {author}
  {\bibfnamefont {B.}~\bibnamefont {Hillebrands}},\ and\ \bibinfo {author}
  {\bibfnamefont {A.~N.}\ \bibnamefont {Slavin}},\ }\href@noop {} {\bibfield
  {journal} {\bibinfo  {journal} {Phys. Rev. Lett.}\ }\textbf {\bibinfo
  {volume} {93}},\ \bibinfo {pages} {047201} (\bibinfo {year}
  {2004})}\BibitemShut {NoStop}%
\bibitem [{\citenamefont {Vansteenkiste}\ \emph {et~al.}(2014)\citenamefont
  {Vansteenkiste}, \citenamefont {Leliaert}, \citenamefont {Dvornik},
  \citenamefont {Helsen}, \citenamefont {Garcia-Sanchez},\ and\ \citenamefont
  {Waeyenberge}}]{Vansteenkiste14}%
  \BibitemOpen
  \bibfield  {author} {\bibinfo {author} {\bibfnamefont {A.}~\bibnamefont
  {Vansteenkiste}}, \bibinfo {author} {\bibfnamefont {J.}~\bibnamefont
  {Leliaert}}, \bibinfo {author} {\bibfnamefont {M.}~\bibnamefont {Dvornik}},
  \bibinfo {author} {\bibfnamefont {M.}~\bibnamefont {Helsen}}, \bibinfo
  {author} {\bibfnamefont {F.}~\bibnamefont {Garcia-Sanchez}},\ and\ \bibinfo
  {author} {\bibfnamefont {B.~V.}\ \bibnamefont {Waeyenberge}},\ }\href@noop {}
  {\bibfield  {journal} {\bibinfo  {journal} {AIP Adv.}\ }\textbf {\bibinfo
  {volume} {4}},\ \bibinfo {pages} {107133} (\bibinfo {year}
  {2014})}\BibitemShut {NoStop}%
\bibitem [{\citenamefont {Chekhov}\ \emph {et~al.}(2018)\citenamefont
  {Chekhov}, \citenamefont {Stognij}, \citenamefont {Satoh}, \citenamefont
  {Murzina}, \citenamefont {Razdolski},\ and\ \citenamefont
  {Stupakiewicz}}]{Chekhov18}%
  \BibitemOpen
  \bibfield  {author} {\bibinfo {author} {\bibfnamefont {A.~L.}\ \bibnamefont
  {Chekhov}}, \bibinfo {author} {\bibfnamefont {A.~I.}\ \bibnamefont
  {Stognij}}, \bibinfo {author} {\bibfnamefont {T.}~\bibnamefont {Satoh}},
  \bibinfo {author} {\bibfnamefont {T.~V.}\ \bibnamefont {Murzina}}, \bibinfo
  {author} {\bibfnamefont {I.}~\bibnamefont {Razdolski}},\ and\ \bibinfo
  {author} {\bibfnamefont {A.}~\bibnamefont {Stupakiewicz}},\ }\href@noop {}
  {\bibfield  {journal} {\bibinfo  {journal} {Nano Lett.}\ }\textbf {\bibinfo
  {volume} {18}},\ \bibinfo {pages} {2970} (\bibinfo {year}
  {2018})}\BibitemShut {NoStop}%
\bibitem [{\citenamefont {Matsumoto}\ \emph {et~al.}(2018)\citenamefont
  {Matsumoto}, \citenamefont {Br\"acher}, \citenamefont {Pirro}, \citenamefont
  {Fischer}, \citenamefont {Bozhko}, \citenamefont {Geilen}, \citenamefont
  {Heussner}, \citenamefont {Meyer}, \citenamefont {Hillebrands},\ and\
  \citenamefont {Satoh}}]{Matsumoto18}%
  \BibitemOpen
  \bibfield  {author} {\bibinfo {author} {\bibfnamefont {K.}~\bibnamefont
  {Matsumoto}}, \bibinfo {author} {\bibfnamefont {T.}~\bibnamefont
  {Br\"acher}}, \bibinfo {author} {\bibfnamefont {P.}~\bibnamefont {Pirro}},
  \bibinfo {author} {\bibfnamefont {T.}~\bibnamefont {Fischer}}, \bibinfo
  {author} {\bibfnamefont {D.}~\bibnamefont {Bozhko}}, \bibinfo {author}
  {\bibfnamefont {M.}~\bibnamefont {Geilen}}, \bibinfo {author} {\bibfnamefont
  {F.}~\bibnamefont {Heussner}}, \bibinfo {author} {\bibfnamefont
  {T.}~\bibnamefont {Meyer}}, \bibinfo {author} {\bibfnamefont
  {B.}~\bibnamefont {Hillebrands}},\ and\ \bibinfo {author} {\bibfnamefont
  {T.}~\bibnamefont {Satoh}},\ }\href@noop {} {\bibfield  {journal} {\bibinfo
  {journal} {Jpn. J. Appl. Phys.}\ }\textbf {\bibinfo {volume} {57}},\ \bibinfo
  {pages} {070308} (\bibinfo {year} {2018})}\BibitemShut {NoStop}%
\bibitem [{\citenamefont {Damon}\ and\ \citenamefont
  {Eshbach}(1961)}]{Damon61}%
  \BibitemOpen
  \bibfield  {author} {\bibinfo {author} {\bibfnamefont {R.~W.}\ \bibnamefont
  {Damon}}\ and\ \bibinfo {author} {\bibfnamefont {J.~R.}\ \bibnamefont
  {Eshbach}},\ }\href@noop {} {\bibfield  {journal} {\bibinfo  {journal} {J.
  Phys. Chem. Solids}\ }\textbf {\bibinfo {volume} {19}},\ \bibinfo {pages}
  {308} (\bibinfo {year} {1961})}\BibitemShut {NoStop}%
\bibitem [{SM2()}]{SM20}%
  \BibitemOpen
  \href@noop {} {}\bibinfo {note} {See Supplemental Material for the
  experimental result of the spin-wave transmission through the air gap, where
  the shaded region represents the air gap.}\BibitemShut {Stop}%
\bibitem [{\citenamefont {Makris}\ and\ \citenamefont
  {Psaltis}(2011)}]{Makris11}%
  \BibitemOpen
  \bibfield  {author} {\bibinfo {author} {\bibfnamefont {K.~G.}\ \bibnamefont
  {Makris}}\ and\ \bibinfo {author} {\bibfnamefont {D.}~\bibnamefont
  {Psaltis}},\ }\href@noop {} {\bibfield  {journal} {\bibinfo  {journal} {Opt.
  Commun.}\ }\textbf {\bibinfo {volume} {284}},\ \bibinfo {pages} {1686}
  (\bibinfo {year} {2011})}\BibitemShut {NoStop}%
\bibitem [{\citenamefont {van Tilburg}\ \emph {et~al.}(2017)\citenamefont {van
  Tilburg}, \citenamefont {Buijnsters}, \citenamefont {Fasolino}, \citenamefont
  {{Th. Rasing}},\ and\ \citenamefont {Katsnelson}}]{Tilburg17}%
  \BibitemOpen
  \bibfield  {author} {\bibinfo {author} {\bibfnamefont {L.~J.~A.}\
  \bibnamefont {van Tilburg}}, \bibinfo {author} {\bibfnamefont {F.~J.}\
  \bibnamefont {Buijnsters}}, \bibinfo {author} {\bibfnamefont
  {A.}~\bibnamefont {Fasolino}}, \bibinfo {author} {\bibnamefont {{Th.
  Rasing}}},\ and\ \bibinfo {author} {\bibfnamefont {M.~I.}\ \bibnamefont
  {Katsnelson}},\ }\href@noop {} {\bibfield  {journal} {\bibinfo  {journal}
  {Phys. Rev. B}\ }\textbf {\bibinfo {volume} {96}},\ \bibinfo {pages} {054437}
  (\bibinfo {year} {2017})}\BibitemShut {NoStop}%
\end{thebibliography}%
\addcontentsline{toc}{chapter}{\bibname}
\end{document}